\documentstyle[12pt]{article} \textwidth=17cm \textheight=23cm
\begin{document}

\centerline {\large\bf What is mass?}

{\bf R. I. Khrapko} \footnote{Moscow Aviation Institute, 4
Volokolamskoe Shosse, 125871, Moscow, Russia.  \\ E-mail:
tahir@k804.mainet.msk.su \quad Subject: Khrapko}

\begin{abstract} Does the mass of bodies depend on their velocity? Is
the mass additive if separate bodies are joined together to form a
composite system? Is the mass of an isolated system conserved?
Different teachers of physics and specialists give different answers to
these questions because there is no general agreement on the definition
of mass..We shall show that the notion of the velocity-dependent
relativistic mass should be given preference over that of the rest
mass. \end{abstract}

\centerline{\bf 1. Introduction}

One of the achievements of the special theory of relativity is the
statement about the equivalence of mass and energy in a sense that the
mass of a body increases with its energy including kinetic energy;
therefore, the mass depends on the velocity of the body. This
relationship is unambiguously interpreted in the works of renowned
physicists.

Max Born (1962): ``The mass of one and the same body is a relative
quantity. It is to have different values according to the system of
reference from which it is measured, or, if measured from a definite
system of reference, according to the velocity of the moving body. It
is impossible that mass is a constant quantity peculiar to each body.''
[1].

Richard Feynman (1965): ``Because of the relation of mass and energy
the energy associated with the motion appears as an extra mass, so
things get heavier when they move. Newton believed that this was not
the case, and that the masses stayed constant.'' [2].

Statements to the same effect can be also found in textbooks.

S. P. Strelkov (1975): ``The dependence of mass on velocity is a
principal proposition of Einstein's mechanics.'' [3].

However, recently there had been a return to the Newton's belief.
According to this belief the mass of a body does not change with
increasing velocity and remains equal to the rest mass. L B Okun' is a
dedicated mouthpiece of this tendency [4, 5]. Earlier, a similar
viewpoint was advocated in the book [6).

L. B. Okun' (1989): ``The mass that increases with speed -- that was
truly incomprehensible. The mass of a body $m$ does not change when it
is in motion and, apart from the factor $c$, is equal to the energy
contained in the body at rest. The mass $m$ does not depend on the
reference frame. At the end of the twentieth century one should bid
farewell to the concept of mass dependent on velocity. This is an
absolutely simple matter!'' [4].

J. Wheeler at al. (1966): ``The concept of relativistic mass is subject
to misunderstanding'' ([6], p. 137).

This opinion is shared by the authors of certain textbooks for
university students published abroad.

R. Resnick et al. (1992): ``\,`The Concept of Mass' by Lev B. Okun (see
Ref. [5] of this letter) summarizes the views held by many physicists
and adopted for use in this book. But there is not universal agreement
on the interpretation of Eq. $$E_0 = mc^2. \eqno (35)$$ This equation
tells us that a particle of mass  $m$ has associated with it a rest
energy  $E_0$. Nevertheless Eq. 35 asserts that energy has mass'' [7]

A serious confusion that  arose from  the reversion  to the Newtonian
concept  of  mass  is  reflected  in   the  following dialogue:

``Schoolboy: `Does mass really depend on velocity, dad?' Father
physicist: `No! Well, yes... Actually, no, but don't tell your
teacher.' The next day the son dropped physics.'' [8].

We hope that we shall succeed in this letter to formulate a rational
approach to the definition of mass.  \medskip

\centerline{\bf 2. A splitting of a definition of mass}

There are two different definitions  of the  inertial mass, coincident
in the non-relativistic context.

Definition  1.   ``In  ordinary   language  the   word  {\it mass}
denotes  something  like  amount  of  substance.  The concept of
substance is considered self-evident.'' (See [1] p.  33.) More
precisely: mass is  defined ``as a  number attached  to each particle
or  body  obtained  by  comparison  with   a  standard body whose mass
is define as unity'' [9].

Definition 2. Mass is a measure of the  inertia of  a body, i.e. the
coefficient of proportionality in the formula $$\hbox{\bf F} =
m\hbox{\bf a}. \eqno (1)$$ or in the formula $$\hbox{\bf p} = m\hbox{\bf
v}. \eqno (2)$$

Because {\bf F}, {\bf a}, {\bf p} and {\bf v} have indisputable
operational definitions, formulas (1) and (2) give the operational
definition of mass. These formulas will be used to make the
aforementioned comparison (see Def. 1) in order to obtain the number
$m$ attached to a body.

For the operational definition of momentum, see [10]. Here is  an
extract from this work: "The meaning of the operational definition
consists  in the identification of  two terms: `definition'  and
`determination'.  The operation  used to  define a momentum  is
essentially  as  follows.  When  a  certain  obstacle  causes a moving
particle to stop, a  force $\hbox{\bf F}(t)$  is measured  with which
the particle acts on  the obstacle  during retardation.  The particle's
initial momentum equals the integral $\hbox{\bf p} = \int\hbox{\bf
F}(t)dt,$ by definition. It is postulated that this integral is
independent of retardation characteristics, i.e. the form  of the
function $\hbox{\bf F}(t).$''

Unfortunately, the attached number determined by formulas (1) and (2)
using the operational definitions of {\bf F}, {\bf a}, {\bf p}, {\bf v}
for one and the same body, i.e. for the same `amount of substance',
turns out to be dependent on the speed of the body; when the body has a
speed, it also depends on the choice of the formula, (1) or (2).
Therefore, the definition of mass for a body in motion splits in three.
`The amount of substance' specified by the attached number from Def. 1
is no longer a measure of a inertia of the moving body.

(a) In order to determine the `amount of substance', i.e. the attached
number from Def. 1, the body must be stopped and formula (1) or (2)
used for a low speed. The number received by this method is called the
rest mass. By definition, this mass does not change when the body
undergoes acceleration.

(b) If the body is not stopped to measure its mass, formula (1) is
known to give no unambiguous result. Because the force and acceleration
are not properties of the body, the coefficient in formula (1) depends
on the direction of the force relative to the body's velocity. As a
matter of fact, this coefficient, in general, becomes a tensor.
Therefore, the definition of the mass by formula (1) is completely
inadequate. It is even not worth considering if the body's speed is not
sufficiently low.

(c) In contrast, formula (2) is valid at any speed including that of
light. For this reason, it and only it gives the operational definition
of the mass of a moving body. Such a mass is a measure of the inertia
of a moving body. It is called the {\it relativistic mass}.

It appears appropriate to cite M. Born once again: ``In physics, as we
must very strongly emphasize, the word {\it mass} has no meaning other
than that given by formula $\hbox{\bf p} = m\hbox{\bf v}$. It is the
measure of resistance of a body to changes of velocity.'' (See [1], p.
33)

At this point, a problem arises. Which of the two masses, the rest mass
of (a) or the relativistic mass of (c), is to be called simply {\it
mass} and denoted by the letter {\it m} without a subscript and thus
regarded as the `chief' mass. This is not a matter of terminology. The
problem has serious psychological and methodological implications.

It can be resolved through the comparison of the properties of
different masses. The rest mass will be denoted by the symbol $m_0$ and
the relativistic mass by the symbol $m$ (otherwise, the latter will
have no simple designation at all).  \medskip

\centerline{\bf 3. System of particles}

If two particles having momenta $\hbox{\bf p}_1 = m_1\hbox{\bf v}_1 $
and $\hbox{\bf p}_2 = m_2\hbox{\bf v}_2 $ join together into a single
whole system, the momenta are known to add up so that $\hbox{\bf p} =
\hbox{\bf p}_1 + \hbox{\bf p}_2. $ Moreover, the four-dimensional
momenta are also summed giving $\hbox{\bf P} = \hbox{\bf P}_1 +
\hbox{\bf P}_2. $ The 4-momentum $\hbox{\bf P}$ is by definition
tangential to the world line of a particle in Minkowski space and its
spatial component equals an ordinary momentum {\bf p}. Hence, the time
component is equal to the relativistic mass $m$: $$\hbox{\bf P} = \{ m,
\hbox{\bf p}\}. $$

This immediately leads to the conclusion that the relativistic masses
are simply summed up: $m=m_1+m_2,$ when particles join together into a
system.

Things  differ  when  rest  masses  come  into question.  In the
4-dimensional  sense, the  rest mass  of a  particle is  the modulus of
its 4-momentum (to an accuracy of c):
$$m_0=\sqrt{m^2-p^2/c^2}.$$
Therefore,  the  rest  mass  of a  pair of  bodies with  rest masses
$m_{01},\quad m_{02}$ is  not  equal  to  the  sum $m_{01}+m_{02}$ but
is determined by   a   complicated   expression   dependent   on
momenta $\hbox{\bf p}_1,\quad\hbox{\bf p}_2$ [4]:
$$m_0 = \sqrt{\left(\sqrt{m_{01}^2+
p_1^2/c^2}+\sqrt{m_{02}^2+p_2^2/c^2}\right)^2-
(\hbox{\bf p}_1+\hbox{\bf p}_2)^2/c^2}.\eqno(3)$$
A similar formula for the rest mass is presented in [6] (c = 1):
$$M^2=(E_{\rm system})^2-(p^x_{\rm system})^2-(p^y_{\rm system})^2-
(p^z_{\rm system})^2.\eqno(4)$$

It follows from formulas (3) and (4) that the rest mass is lacking the
property of additivity. We think that physicists do not mean the rest
mass when they speak about beauty as a criterion for truth.  \medskip

\centerline{\bf 4. A violence to mind}

The thing is that both the relativistic mass (a time component of
4-momentum) and the rest mass (its modulus) obey the conservation law.
This is ascertained in [4].

However, it is not so simple to accept that a non-additive quantity is
conserved. Indeed, according to (3) and (4), the rest mass of a system
does not change as a result of particle collisions or nuclear
reactions. But as soon as a system of two moving bodies is mentally
divided into two separate bodies, the rest mass will change because the
rest mass of the pair is not equal to the sum of the rest mass of the
bodies of the pair. In our opinion, the use of non-additive notions
entails a serious intellectual burden: a pair of photons, each having
no rest mass, does have a rest mass.

Another very difficult question is: ``Does energy have a rest mass?''
The correct answer may be as follows: the energy of two photons will
have a rest mass when they move in opposite directions. A system of two
photons will have zero rest mass if they move in the same direction
[4]. Thus, it appears that even the authors of the textbook [7] failed
to solve the problem.

Furthermore, photons moving in the same direction have no rest mass
while the rest mass of the body which emitted them decreases.
Therefore, it may be suggested that some of the body's rest mass has
been converted into the massless energy of photons. However, according
to (3), (4) the rest mass of the system body-photons has been conserved
during radiation!

Unable to bear such an intellectual burden, the advocates of the rest
mass concept refuse to adopt the law of conservation of the rest mass
of a system, in defiance of the formulas (3), (4). Now, they state that
``rest mass of final system increases in an inelastic encounter'' ([6],
p. 121). In contrast, nuclear reactions lead to `the mass defect'. For
example, in the synthesis of deuteron, p + n = D + 0.2 MeV, its rest
mass is less than that of the neutron and proton.

At the same time, it follows from formulas (3), (4) that there must be
no rest mass `defect' during nuclear reactions. In our example, the
allegedly lacking rest mass of the system at stage D + 0.2 MeV is
actually provided by a massless $\gamma$-quantum with the energy of 0.2
MeV. This disturbs the additivity of the system's rest mass.

It is easy to understand why the schoolboy dropped physics in the face
of such a confusion concerning the rest mass.  \medskip

\centerline{\bf 5. Underlying psychologic reason}

For all that, many physicists consider the rest mass to be the `chief'
one and denote it by the symbol $m$ instead of $m_0.$ Simultaneously,
they discriminate against the relativistic mass and leave it without
notation. This causes an additional confusion  making  it  sometimes
difficult  to   understand  which mass  is  really  meant.  This
situation  is exemplified  by the statement from [7] cited above.

These physicists agree that the mass of a gas in a state of rest
increases upon heating  because the  energy contained  in it grows.
However,  there  seems to  exist a  psychological barrier which
prevents relating this rise to a larger mass  of individual molecules
due to their high thermal velocity.

The said physicists sacrifice the  concept of  a mass  as a measure  of
inertia, sacrifice the additivity of mass and the equivalence  of mass
and energy to a  label attached to each particle and bearing
information  about  a  constant  `amount  of substance', just because
such a label is  in line  with the  deeply ingrained Newtonian
concept   of   mass.   For   them,   radiation   that "transmits
inertia between emitting and absorbing bodies'' (according  to  A
Einstein [11]) has no mass.

The  main  psychological  problem is  how to  establish the identity
between  mass  and  energy  (which  varies)  and regard these two
entities as one. It is easy to accept that $E_0=m_0c^2$ for a body at
rest. The authors of  Ref. [6]  entitled Chapter  13 as ``The
equivalence  of  energy  and  rest  mass"\footnote{The title is
characteristically ambiguous implying the equivalence between the {\it
rest} energy and the rest mass.} It   is  more difficult  to admit
that the  formula $E  = mc^2$  is valid  for any speed.   The
remarkable formula   $E=mc^2$ is    described   by L. B. Okun' as
`ugly' [4].

Transition from the rest mass to the relativistic one  in the
relativistic theory  appears to  encounter the  same psychological
problems as transition from proper to relative time.

     It is appropriate to quote from Max Planck here:
\begin{quote}
``An important scientific innovation rarely makes its way by gradually
winning over and converting its opponents: it rarely happens that Saul
becomes Paul. What does happen is that its opponents gradually die out
and that the growing generation is familiarized with the idea from the
beginning: another instance of the fact that the future lies with
youth. For this reason a suitable planning of school teaching is one of
the most important conditions of progress in science." [12].
\end{quote}

     Unfortunately, the important concept of relativistic mass is
carefully isolated from youth: the present paper has been rejected by
editors of the following journals: ``Russian Physics Journal", ``Kvant"
(Moscow), ``American Journal of Physics", ``Physics Education" (Bristol).
``Physics Today".

\centerline{\bf 6. Conclusions}

Thus,  the  relativistic  mass  has  a  natural operational definition
based  on  the  formula $\hbox{\bf p}  = m\hbox{\bf v}.$ It is
additive and obeys the law  of conservation. Also, it  is equivalent
to both energy  and  gravitational  mass.  It  should  be referred  to
as mass and denoted by the letter $m$.

The rest mass  is not  conserved or  lacks the  property of additivity.
Here, the advocates of the rest mass concept contradict themselves; at
first, they justly maintain that the rest mass is conserved but not
additive, then they say that it is additive but not conserved.
It  is  not  equivalent  to  energy.  It  should be denoted as $m_0$
and used  with caution  especially if  the notion is applied to a
system of bodies.

The relativistic mass together with momentum are transformed  as
coordinates  of  an  event during  transition to a new inertial
laboratory:
$$m=\frac{m'+p'v/c^2}{\sqrt{1 - v^2/c^2}},
\quad p=\frac{p'+m'v}{\sqrt{1 - v^2/c^2}}.$$
Specifically, if $p' = 0$ then $m' = m_0,$ and
$$m=\frac{m_0}{\sqrt{1 - v^2/c^2}}, \quad
p=\frac{m_0v}{\sqrt{1 - v^2/c^2}}.$$

It is worthwhile to note in conclusion that if instead of the
coordinates $t, x,$...  we use  the coordinates  $t', x',$... the
relativistic  mass $ m$  and  the  rest  mass  $m_0$, which  are both
scalars, will be expressed by the formulas
$$mc=u^{i'}p^{j'}g_{i'j'},\quad m_0c=\sqrt{p^{i'}p^{j'}g_{i'j'}},$$
which  are  valid  for  the  curved  space  of  GTR.   Here,  $u^{i'},
p^{j'}$ and $g_{i'j'}$ are  the  unit  vector  of  the  experimentalist,
4-momentum  of the   body,   and   metric   tensor   of   the  new
coordinates, respectively.  It  is  assumed  that  for  the  initial
coordinates $t,  x,$..., $u^i=\delta_0^i, g_{00}= 1, g_{ii}= -1,$...

A  photon  has  no  rest  mass-energy,  hence  no  proper frequency.
But  its  mass-energy  and  frequency   can  be measured in experiment
as $E = h\nu = cu^ip^jg_{ij}$ and prove to be of any value depending on
the experimenter's speed. I thank  G. S. Lapidus whose  comments
helped  to improve the text of this paper. \medskip

\noindent This paper has been published in {\it Physics - Uspekhi} {\bf 43}
(12) 1267 (2000), {\it http://www.ufn.ru}\\
{\it Uspekhi Fizicheskikh Nauk} {\bf 170} (12) 1363 (2000), {\it
http://www.ufn.ru}\\
{\it http://www.mai.ru/projects/mai\_works/index.htm}

This topic is elaborated in {\it physics/0103008}. \medskip

\centerline{\bf References }

1.  Born M.,  Einstein's Theory  of Relativity  (New York:  Dover
Publ., 1962) p. 269.

2.  Feynman R., Character  of Physical  Law (London:  Cox and Wyman,
1965) p. 76.

3. Strelkov S. P., Mechanics (Moscow: Nauka, 1975) p. 533 (in Russian).

4. Okun' L. B., ``The concept of mass'', Soviet Physics Uspekhi {\bf
32} (7), 629--638 (1989).

5. Okun' L. B., ``The concept of mass", Physics Today {\bf 42} (6), 31
(1989)

6.  Taylor E.  F.,  Wheeler  J. A.,  Spacetime Physics (San Francisco:
W.H. Freeman, 1966).

7. Resnick R., Halliday D., Krane K. S.,  Physics, Vol. 1 (New  York
Wiley, 1992), p. 166, 167.

8. Adler C. G. ``Does mass really depend on velocity, dad!" Am. J.
Phys. {\bf 55}, 739 (1987).

9.  Alonso M.,  Finn E. J.,  Physics  (Wokingham, England:
Addison- Wesley, 1992) p. 96.

10.  Khrapko R. I.,  Spirin G. G., Ramrenov V. M.,  Mechanics (Moscow:
Izd. MAI, 1993).

11. Einstein A. ``Ist die Tragheit eines Korpers von seinem
Energiegehalt abhangig?' Ann. d. Phys.  {\bf 18},  639 (1905).

12. Planck M., The Philosophy of Physics (George Allen \& Unwin Ltd,
    London, 1936), p. 90.
\end{document}